\documentclass{aa}
\usepackage{graphicx}
\usepackage{txfonts}
\usepackage{natbib}            
\bibpunct{(}{)}{;}{a}{}{,}     

\titlerunning{Simulations to optimize the $DM_2$ conjugation altitude in an MCAO system}
\authorrunning{B. Femen\'{\i}a \& N. Devaney}

\begin{document}

\title{Optimization with numerical simulations of the conjugate altitudes
of deformable mirrors in an MCAO system }

\author{B. Femen\'{\i}a and N. Devaney}

\institute{GTC Project, Instituto de Astrof\'{\i}sica de Canarias, V\'{\i}a L\'actea s/n,
38200 La Laguna, SPAIN\\
\email{bfemenia@ll.iac.es, ndevaney@ll.iac.es}}

\offprints{B. Femen\'{\i}a}

\date{Received 12 February 2003/ Accepted 6 April 2003}

\abstract{This article reports on the results of simulations conducted to assess
the performance of a modal Multi-Conjugate Adaptive Optics (MCAO)
system on a 10m telescope with one Deformable Mirror (DM) conjugated
to the telescope pupil and a second DM conjugated at a certain altitude
above the pupil. The main goal of these simulations is to study the
dependence of MCAO performance upon the altitude of the high-altitude
conjugated DM and thereby determine its optimal conjugation. The performance
is also studied with respect to the geometry of the Guide Star constellation
when using constellations of Natural Guide Stars (NGS), which are
rare, or constellations of Laser Guide Stars (LGS) which would allow
large sky coverage.
}

\maketitle

\section{Introduction\label{sec:intro}}

Adaptive Optics (AO) was first proposed by \citet{BabcockH:posca} as a means to
compensate for the loss of spatial resolution of telescopes due to the
disturbing effects caused by atmospheric turbulence \citep{RoddierF:effato} and
thereby obtain diffraction limited resolution. A major limitation of AO is the
small size of the Field of View (FoV) over which the correction is effective,
referred to as the isoplanatic patch. Under good seeing conditions a typical
value for the isoplanatic radius is just a few (2-3) arcsec in the visible
($0.5\mu m$) and about 12-18 arcsec in the near infrared ($2.2\mu m$). The
concept of MCAO as a technique to deliver large corrected FoVs was proposed by
\citet{DickeR:phacd} and later developed by \citet{BeckersJ:incsi} with the main
goal of increasing the corrected FoV. It did not however receive much attention
at that time as the AO community focused on conceptual problems such as
obtaining adequate sky coverage with and without the use of Laser Guide Stars
(LGS) \citep{SandlerD:adaod,LeLouarnM:lgs38} and the problems associated with
LGS-based AO systems, in particular tip-tilt indetermination and the cone
effect. During the early '90s the concept of atmospheric turbulence
reconstruction in 3-D (i.e.  tomography) was proposed as a solution to overcome
the problems associated with sky coverage and the cone effect in LGS-based AO
systems \citep{TallonM:adatl}. Shortly afterwards, the theoretical foundations
of MCAO were developed \citep[see e.g.][]{JohnstonD:anama,EllerbroekB:fisop}.
In the last few years it has been recognised that tomography and MCAO offer the
possibility of attaining large corrected FoVs and there is a large on-going
effort in MCAO on both the theoretical side \citep[see
e.g.][]{RagazzoniR:modta,FuscoT:isoao,RagazzoniR:ada100,RigautF:prilp,FuscoT:optwr}
and using numerical simulation \citep[see
e.g.][]{FlickerR:comma,CarbilletM:clolp,ConanJ:mulao,BelloD:perma,EllerbroekB:wavop,TordiM:simlo,TokovininA:Optmt,FemeniaB:numsm,EllerbroekB:metct,LeLouarnM:mulao}
to assess the performance of MCAO systems.

Conventional AO systems employ a single Deformable Mirror (DM) usually
conjugated to the telescope entrance pupil. MCAO requires at least
a second DM conjugated  to some altitude above the pupil. The addition
of a second DM represents a modest increase in complexity (of course
the wave-front sensor also has to be modified). Such a dual-conjugate
mode is considered as a short term upgrade to the AO system being
designed for the 10m GTC telescope \citep{DevaneyN:adawb} at the
Observatorio del Roque de los Muchachos (ORM, Spain). In this work
we present numerical simulations of a dual-mirror modal MCAO system
for the case of a 10-m telescope. The main goal of our simulations
is to determine the optimal conjugation altitude of the high-altitude
DM (from now on referred to as $DM_{2}$) assuming that the low-altitude
DM ($DM_{1}$) is conjugated to ground level. Our numerical simulations
investigate the dependence of the optimal high-altitude conjugation
altitude on the nature of the GS constellation (either NGS or NGS
plus LGS constellations are considered), the geometry and the number
of reference sources in the GS constellation. These simulation results
will be compared with theoretical predictions based on the estimates
of \citet{TokovininA:limpt}  for the optimal conjugation altitude
of $DM_{2}$.

\section{MCAO model}

\subsection{Modal tomography approach\label{sec:maths}}

In modal tomography \citep{LawrenceG:wavtz,RagazzoniR:modta} one seeks to
estimate the coefficients of Zernike expansions of the wavefronts over
metapupils defined at the conjugate altitudes of the DMs. The metapupils are
circles enclosing the footprints of the beams from the guide stars. If
$\overrightarrow{b}_\mathrm{(g)}$ denotes the Zernike polynomial decomposition
of the incoming wavefront when looking at guide star $g$ in the direction
$\overrightarrow{\theta }_\mathrm{(g)}$ (i.e. $W(\overrightarrow{x};\,
\overrightarrow{\theta }_\mathrm{(g)})\, =\sum_{i} \,
b_{i,\mathrm{(g)}}Z_\mathrm{i}(\overrightarrow{x}/R)$, with R the pupil radius)
then the Zernike polynomial decomposition of the wavefront defined on each of
the metapupils to be reconstructed at heights $h_{l}$ ($l=1\ldots L$) is
$\left\{ \overrightarrow{a}^{(l)}\right\} _{l=1\ldots L}$ which is related to
$\left\{ \overrightarrow{b}_{(g)}\right\} _{g=1\ldots G}$ by :

\begin{equation}
\overrightarrow{b}^\mathrm{T}=\mathbf{P}\, \overrightarrow{a}^\mathrm{T},\label{eq:tomo_eq}\end{equation}
 where the subscript $^\mathrm{T}$ denotes transpose and the different terms
in Eq.~(\ref{eq:tomo_eq}) are:

\begin{eqnarray}
\overrightarrow{b} & = & \left[\begin{array}{cccc}
 \overrightarrow{b}_{(1)}\left(\overrightarrow{\theta }_{(1)}\right) & \overrightarrow{b}_{(2)}\left(\overrightarrow{\theta }_{(2)}\right) & \ldots  & \overrightarrow{b}_\mathrm{(G)}\left(\overrightarrow{\theta }_\mathrm{(G)}\right)\end{array}\right],\\
\mathbf{P} & = & \left[\begin{array}{cccc}
 \mathbf{P}(\overrightarrow{\theta }_{(1)};h_{1}) & \mathbf{P}(\overrightarrow{\theta }_{(1)};h_{2}) & \cdots  & \mathbf{P}(\overrightarrow{\theta }_{(1)};h_\mathrm{L})\\
 \mathbf{P}(\overrightarrow{\theta }_{(2)};h_{1}) & \mathbf{P}(\overrightarrow{\theta }_{(2)};h_{2}) & \cdots  & \mathbf{P}(\overrightarrow{\theta }_{(2)};h_\mathrm{L})\\
 \vdots  & \vdots  & \ddots  & \vdots \\
 \mathbf{P}(\overrightarrow{\theta }_\mathrm{(G)};h_{1}) & \mathbf{P}(\overrightarrow{\theta }_\mathrm{(G)};h_{2}) & \cdots  & \mathbf{P}(\overrightarrow{\theta }_\mathrm{(G)};h_\mathrm{L})\end{array}\right],\label{eq:P_matrix}\\
\overrightarrow{a} & = & \left[\begin{array}{cccc}
 \overrightarrow{a}^{(1)} & \overrightarrow{a}^{(2)} & \ldots  & \overrightarrow{a}^{(L)}\end{array}\right].
\end{eqnarray}
 and in Eq.~(\ref{eq:P_matrix}) each of the $\mathbf{P}(\overrightarrow{\theta }_\mathrm{(g)};h_\mathrm{l})$
is a Zernike projection matrix relating the Zernike decomposition
coefficients of the wavefront defined on the metapupil ($\overrightarrow{a}^\mathrm{(l)}$)
and $\overrightarrow{b}_\mathrm{(g)}^\mathrm{(l)}\left(\overrightarrow{\theta }_\mathrm{(g)};h_\mathrm{l}\right)$
are the Zernike decomposition coefficients of the wavefront defined
on the intersection of the metapupil and the propagation path at the
$l$th layer when looking toward $\overrightarrow{\theta }_\mathrm{(g)}$:

\begin{equation}
\overrightarrow{b}_\mathrm{(g)}^\mathrm{(l)}\left(\overrightarrow{\theta }_\mathrm{(g)};h_\mathrm{l}\right)=\mathbf{P}(\overrightarrow{\theta }_\mathrm{(g)};h_\mathrm{l})\, \overrightarrow{a}^\mathrm{(l)}.\label{eq:tomo_eq2}\end{equation}

The existence of such $\mathbf{P}(\overrightarrow{\theta
}_\mathrm{(g)};h_\mathrm{l})$ matrices is demonstrated in
\citet{RagazzoniR:modta} and an analytical expression for their computation is
provided in \citet{FemeniaB:anapz} and depends exclusively on the geometry of
the problem (i.e. on the metapupil radius, radius of the circular region defined
by the intersection of the metapupil and the propagation path and the relative
position of the circular region with respect to the metapupil center). The
required $\left\{ \overrightarrow{a}^{(l)}\right\}_{l=1\ldots L}$ coefficients
may be obtained from Eq.~(\ref{eq:tomo_eq}) by a proper inversion method
(e.g. SVD or least square estimator, optimal reconstructor or maximum-a-priori
reconstructor).

\subsection{Implementation in the simulation\label{sec:math2simu}}

The previous description of modal tomography does not give any indication
as to how the Zernike polynomial coefficients $\overrightarrow{b}$
of the GS wavefronts are retrieved from real data. In a real wavefront
sensor the data will be wavefront local slope measurements (as in
the case of Shack-Hartmann and pyramid wavefront sensors) or wavefront
local curvature values (as in the case of the curvature sensor). We
require a reconstruction process from real measurements to the metapupil
Zernike coefficients $\overrightarrow{a}$ without having to first
reconstruct the GS Zernike coefficients $\overrightarrow{b}$. This
is achieved by means of the \emph{interaction matrix} $\mathbf{M}$
of the MCAO system linking wavefront sensor outputs ($\overrightarrow{s}$)
and metapupil Zernike coefficients: $\overrightarrow{s}=\mathbf{M}\overrightarrow{a}$.
While the $\mathbf{P}$ matrix in Sect.~\ref{sec:maths} is obtained
from purely geometric considerations, building the interaction matrix
$\mathbf{M}$ requires explicit knowledge of the WFS such as the extended
size of the GS spot as seen by the WFS, pixelization effects in the
WFS or the fact that the WFS sees only a limited FoV. This
leads us to use interaction matrices generated by the simulation in
which we sequentially place different Zernike polynomials on each
layer and register the local tilts produced on each subaperture of
each Shack-Hartmann Sensor (SHS) in our MCAO system. In the simulations
using LGSs we consider that there is a global tilt subtraction for
the wavefront coming from each LGS.

\section{Description of the software employed: CAOS\label{sec:CAOS}}

The main software tool used is the IDL-based Code for Adaptive Optics
Systems \citep[CAOS,][]{CarbilletM:caos3,CarbilletM:laos}. The main
feature of CAOS is its modular structure: each elementary physical
process is modeled by a specific set of routines (i.e. a module) ,
e.g. the simulation of atmospheric turbulence or the propagation of
the wavefront through turbulent layers. This modular design makes
CAOS a very versatile tool as it can be extended by the user with
new modules as required. The other main feature of the CAOS software
is its graphic user interface, the {}``Application Builder'' \citep{FiniL:caosab},
which helps to design simulation projects in a very straightforward
way even for users with no knowledge of IDL. After the graphical design,
the Application Builder automatically generates an IDL code which
can be run as the desired simulation. In order to simulate
MCAO systems new modules have been developed in addition to the modules
in the standard CAOS library. A dedicated module conducts the calibration
of the MCAO system yielding the interaction matrix: at each layer
to be reconstructed we place a single polynomial $Z_\mathrm{j}$ at a time
and perform the geometrical propagation (i.e near-field approximation)
of the wavefront error corresponding to this $Z_\mathrm{j}$ along the line
of sight of each Guide Star (GS) down to the SH sensors. The vector
of x- and y-slopes from all the SH sensors gives us the column of
the matrix associated with the j-th Zernike polynomial on the considered
layer; this process is then repeated for all the modes to be reconstructed
on each layer. The reconstruction process is done with a dedicated
module which uses a Least Square Estimator (LSE) and by inverting
with Singular Value Decomposition (SVD) the linear relation $\overrightarrow{s}=\mathbf{M}\overrightarrow{a}$
between the measured slopes ($\overrightarrow{s}$) and the coefficients
($\overrightarrow{a}$) on each reconstructed metapupil. The multi-mirror
reconstructor is simulated assuming perfect Zernike DMs in the sense
that any mirror is able to generate a pure Zernike mode up to a given
mode $j$ and allows different degrees of correction to be used at
different layers.

\section{Simulation description \& simplifications\label{sec:simplifications}}

\subsection{Generation of Guide Stars (GS)}

NGS are simulated as point-like sources at an infinite distance from
the telescope and sampling a cylindrical turbulence volume. LGS are
simulated as extended sources at a finite distance so that they sample
a cone shaped turbulence volume. The width of the LGS is evaluated
taking into account the broadening due to diffraction by a launching
telescope of 0.5~m diameter, the upward propagation of the laser
beam, the perspective elongation of the LGS beacon as seen from the
observing telescope and the geometry of the beacon size averaged over
the range gate (for Rayleigh LGS) or width of the Na-layer (for Na
LGS) using the expression presented in \citet{EllerbroekB:metct}:

\begin{equation}
\theta _\mathrm{B}^{2}=\left(\frac{\lambda }{d_\mathrm{T}}\right)^{2}+\left(\frac{\lambda }{r_{0}}\right)^{2}+\frac{1}{3}\left(\frac{L\, \delta h}{h^{2}}\right)^{2}+\left(\frac{d_\mathrm{T}\, \delta h}{2\sqrt{3}h^{2}}\right)^{2}\label{eq:width_LGS}\end{equation}
where $\lambda $ is the wavelength of observation ($\lambda =589.3\ \mathrm{nm}$
for Na~LGS, for Rayleigh LGS we consider $\lambda =355\ \mathrm{nm}$), $d_\mathrm{T}$
the launching telescope diameter, $r_{0}$ the Fried parameter at
the WFS operation wavelength, $L$ the distance between the launching
telescope and the SHS subaperture, $h$ the altitude of the LGS and
$\delta h$ the {}``range gate''. By {}``range gate'' we mean
the vertical distance of the LGS on which we integrate while it is
propagating upward. Such a value can be fixed by us when considering
Rayleigh LGS (by means of e.g. a Pockels cell). In the case of a Na
LGS it is not possible to fix a range gate since the Sodium layer
drifts in altitude and thus in the above expression we must use the
full width of the Sodium layer weighted by the fact that the sodium
layer has a Gaussian density profile so that an appropriate value
in the above expression is $\delta h=5$~Km. Then the LGSs are projected
onto the WFS so that they look like circular Gaussian spots
of width equal to $\theta_\mathrm{B}$. This neglects the differential elongation
due to the finite distance from the telescope to the LGS but on the
other hand such an approach is pessimistic since the LGS is not circular
but elongated and in practice this involves a loss of resolution only
along the elongation direction.

\subsection{Generation of atmospheric turbulence\label{sec:simp_atm}}

For the purpose of our simulations we have constructed a 7-layer profile
summarized in Table~\ref{tab:averballprofile}. This profile was
built using the measurements of balloon flights \citep{VerninJ:optsp}
launched from the Observatorio del Roque de los Muchachos (ORM). Data
from the 6 available balloon flights were used to determine an averaged
profile in the following way; first all profiles from the balloon
measurements were normalized to have the same total $r_{0}$ as the
annual mean (i.e. $r_{0}(500\mu \mathrm{m})=0.15\, \mathrm{m}$) at
ORM and then a continuous average profile was generated. Then, following~\citet{EllerbroekB:adapp},
a 7-layer turbulence profile was obtained by performing a discrete
fit to the continuous average profile that matches the first 14 moments
of the continuous turbulence profile. 

\begin{table}

\caption{The ORM average balloon 7-layer profile. \label{tab:averballprofile}}

\begin{tabular}{ccccc}
\hline 
Height$^{\mathrm{a}}$&
$C_\textrm{n,i}^{2}$ &
$C_\textrm{n,i}^{2}$ &
$r_{0}(0.5\mu \textrm{m})$&
$r_{0}(2.2\mu \textrm{m})$\\
(m)&
($m^{1/3}$ )&
(\%)&
(m) &
(m)\\
\hline
4.5&
$8.08\times 10^{-14}$&
$23.0$ &
$0.363$ &
$2.15$\\
463&
$1.04\times 10^{-13}$&
$29.6$ &
$0.312$ &
$1.85$\\
1483&
$8.64\times 10^{-14}$&
$24.6$ &
$0.348$ &
$2.06$\\
4840&
$3.86\times 10^{-14}$ &
$11.0$ &
$0.565$ &
$3.34$\\
11122&
$2.25\times 10^{-14}$&
$6.4$ &
$0.781$ &
$4.62$\\
14906&
$1.76\times 10^{-14}$&
$5.0$ &
$0.905$ &
$5.36$\\
18635&
$1.40\times 10^{-15}$&
$0.4$&
$4.13$&
$24.4$\\
\hline
\end{tabular}

\begin{list}{}{} \item[$^{\mathrm{a}}$] Height refers to altitude above the telescope. \end{list}
\end{table}

Atmospheric turbulent phase screen generation was performed by summing
a finite set of Zernike polynomials weighted by a set of coefficients
drawn from a Gaussian multivariate distribution of zero mean and covariance
matrix as given in \citet{NollR:zerpa}, i.e. assuming the turbulence
follows Kolmogorov statistics. The truncation of the infinite
series implies missing high spatial frequency components. This is
easily obviated by truncating at a large enough mode $j_{\mathrm{max}}$.
The total Strehl ratio due to all missing modes on all layers is estimated
to amount to about 10\% and this reduction is applied to the iso-Strehl
maps. A marginal effect is due to the actual statistics followed by
the Zernike coefficients on portions of the phase screens the same
size as the telescope pupil. This problem can also be circumvented
by the use of a large enough number of Zernike modes when generating
the turbulence phase screens.

\subsection{Propagation through turbulence and wavefront sensing\label{sec:simp_wfs}}

Propagation of light through atmospheric turbulence is performed by
adding linearly the phase fluctuation produced by each turbulent layer
which assumes the near-field approximation \citep{RoddierF:effato}
and corresponds to neglecting diffraction effects in the propagation
of the phase of the complex field between layers and from layers to
the ground (a similar approximation is implicit when assuming thin
layers in Sect.~\ref{sec:simp_atm}: diffraction effects between
the top and the bottom of the layer are neglected). \citet{EllerbroekB:wavop}
presents results using a code implementing wave-optics for phase propagation
and makes a comparison of the results obtained under the near-field
approximation for the case of an MCAO system at the Gemini telescope.
He concludes that including Fresnel-diffraction in the propagation
of light involves a phase rms of only a few nm (<20~nm) with respect
to the results obtained with the geometric propagation. Another effect
related to not including Fresnel diffraction is that we neglect the
effect of scintillation on wavefront sensing. CAOS implements a wave-optics
code to simulate the WFS. However the problems of scintillation are
not handled since they do not appear in the near-field approximation.
Our simulations assume Shack-Hartmann WFS for which the performance
dependence on the scintillation has been studied in \citet{VoitsekhovichV:infsp}
in the case of open-loop measurements. A direct extrapolation of the
results of \citet{VoitsekhovichV:infsp} indicates that for our case
with $r_{0}(500\, \mathrm{nm})=0.15$~m scintillation effects in
the WFS will involve errors $<$ 8\% in the estimation of the high order
reconstructed modes. A much more important simplification is that
of no noise in the WFS procedure as we are not concerned at this stage
with problems associated with sky coverage and by not considering
noise we are in the limit of very bright objects.

\subsection{Least Square Estimator (LSE) reconstructor\label{sec:simp_LSE}}

Several works \citep{FuscoT:phael,BrusaG:multAO,FlickerR:comma,FuscoT:optwr,ConanJ:mulao}
have focused on estimators for MCAO which assume a real-time knowledge
of the statistics of phase perturbations and the turbulence vertical
distribution to derive optimal reconstructors operating in open-loop:
the so-called Maximum \textit{a posteriori} (MAP) estimators. In conditions
of low Signal-to-Noise Ratio (SNR) it has been shown \citep{FlickerR:comma,ConanJ:mulao,LeLouarnM:mulao}
that these optimal reconstructors perform better than the SVD-based
reconstruction while yielding similar results in the case of high
SNR values. Since we work in the limit of very bright objects (see
Sect.~\ref{sec:simp_wfs}) nothing justifies using MAP estimators
here except for the advantage of including knowledge of the vertical
distribution of turbulence. On the other hand, it is clear that a
MCAO system will be operated in closed-loop which imposes no problem
for the implementation of the LSE reconstructor while, to our knowledge,
statistics of the residual signal on the WFS have not been derived
analytically to be implemented in the design of the MAP reconstructor.

\subsection{Perfect Zernike deformable mirrors\label{sec:simp_DM}}

The DMs considered in these simulations are assumed to be able to
reproduce exactly the Zernike modes being reconstructed, that is,
if the command is that of generating $a_\mathrm{j}Z_\mathrm{j}$ (with $\mathrm{j=2,\ldots ,153}$)
on $DM_{1}$ then this is exactly what we get (idem with $DM_{2}$
but $\mathrm{j=2,\ldots ,91}$) . This assumption neglects the influence functions
of real DMs which in addition to the commanded $Z_\mathrm{j}$ mode will
probably introduce other modes at a much lower level. Realistic influence
functions will be considered in future work using more detailed numerical
simulations.

\subsection{Open-loop operation.\label{sec:simp_OL}}

Our simulation assumes open-loop operation. In doing this we can easily
perform a statistical analysis by generating independent input turbulent
wavefronts and retrieving the response of the MCAO system. Open-loop
operation neglects degradation of the MCAO system performance due
to time-delay errors while on the other hand it gives an underestimation
of the performance of the system as compared to closed-loop operation
since the WFS is more affected by problems due to the restricted FoV
in the SHS and the extended size of the GS. 

\begin{figure}
  \resizebox{\hsize}{!}{\includegraphics{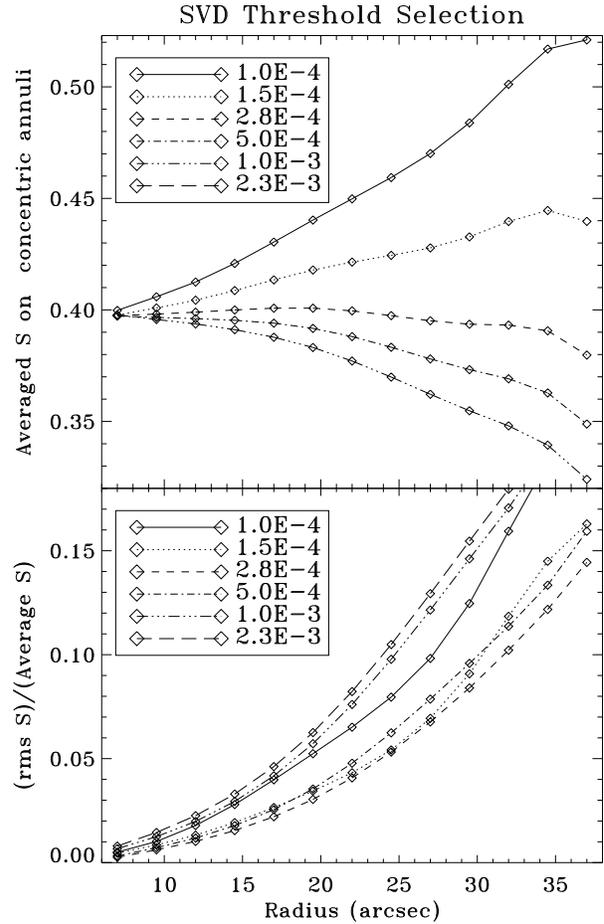}}
  \caption{Example of SVD threshold selection for the case of 3NGS on a
circumference of $45\arcsec $ with $DM_{2}$ conjugated to 4.7 Km.The normalized
S rms curve corresponding to $thr_{2}=1.5\times 10^{-4}$ is slightly displaced
upward so as not to be superimposed on the curve corresponding to
$thr_{3}=2.8\times 10^{-4}$. Here we selected the $thr_{3}$ curve as the one
with the best compromise between the highest S ratio value and the lowest S
$rms$.}
  \label{fig:SVD_selection}
\end{figure}

\section{Simulation results\label{sec:results}}

\begin{figure*}
\centering
  \includegraphics[width=17cm]{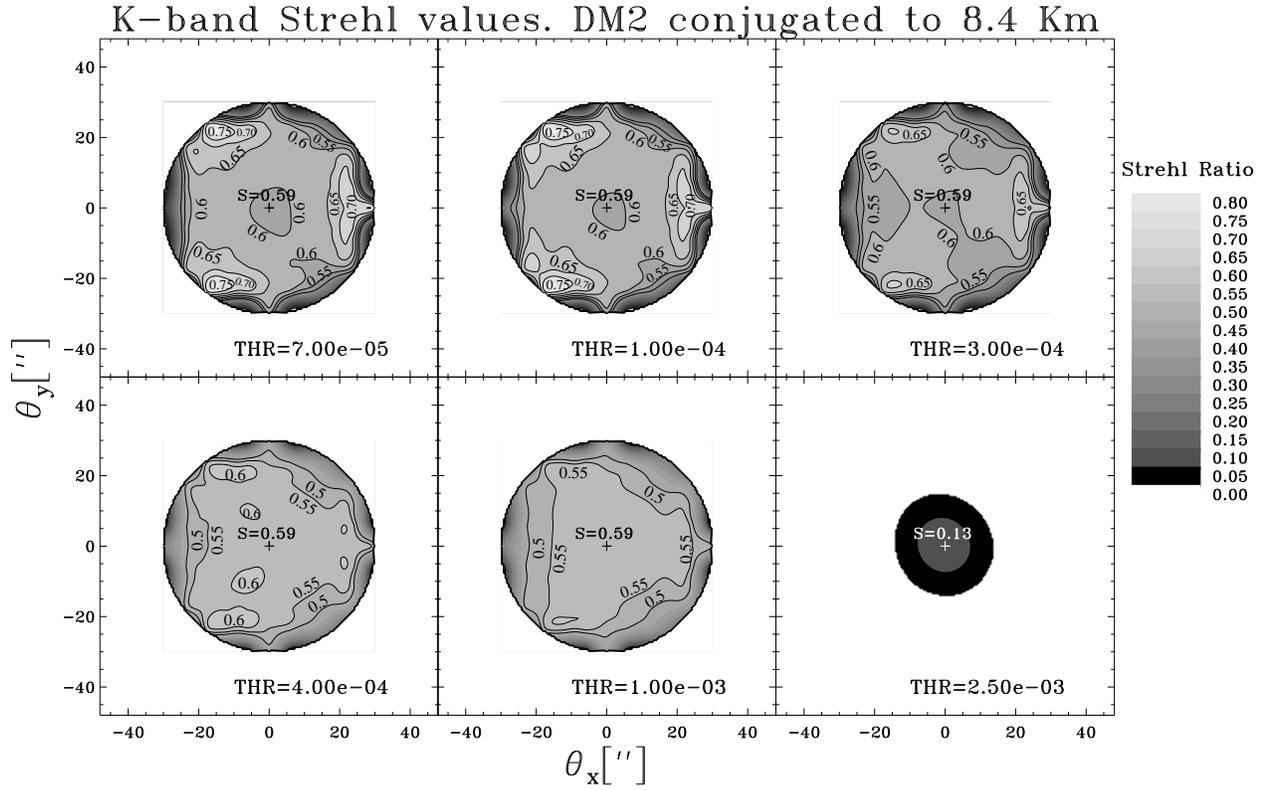}
  \caption{Strehl maps for several SVD threshold values in a reconstructed FoV
           of $1\arcmin $ diameter with $DM_{2}$ conjugated to 8.4~Km and 3NGS
           on a circumference of $30\arcsec $ radius.}
  \label{fig:iso_3NGS_ABP_30arc_DM2_08Km}
\end{figure*}

In the case of single-conjugate AO with moderate-good correction,
the performance may be characterised adequately by a single parameter,
namely the on-axis Strehl ratio. The Strehl ratio can be maximised
by choosing a threshold value in the SVD reconstruction; singular
values lower that the threshold are set to zero and this eliminates
modes to which the system is insensitive. In the case of an MCAO system,
the performance to be determined is the Strehl ratio over the field
of view. We find in our simulations that the Strehl ratio maps are
sensitive to the SVD threshold (for example, see Fig.~\ref{fig:iso_3NGS_ABP_30arc_DM2_08Km}).
Since the SVD threshold cannot be selected \emph{a priori} we need
criteria for selecting the optimal threshold values for each configuration.
In order to characterize SR maps, we extract two parameters which
quantify the average Strehl ratio over the field and the amount of
variation of the Strehl ratio over the field. For a given Strehl map
we determine the value of the Strehl ratio averaged on concentric
annular radii (the first annulus is the inner circle of radius $7\arcsec $
and concentric annuli have an outer radius $2.5\arcsec $ larger than
the inner radius) and the root mean square deviation of Strehl in
each annulus with respect to the averaged Strehl ratio on the circle
of radius equal to the annulus outer radius. As shown in the example
in Fig.~\ref{fig:SVD_selection} we obtain two families of curves:
the first set of curves shows for each SVD threshold value the evolution
of Strehl averaged on annuli as a function of the annulus outer radius
while the second family of curves shows the evolution of Strehl $rms$.
This process is conducted on contour plots such as those depicted
in Fig.~\ref{fig:iso_3NGS_ABP_30arc_DM2_08Km}. To produce these
contour plots we interpolate the reconstructed FoV sampled at 15x15
locations in the simulation to a finer grid. Due to this interpolation
when we get close to the edges we observe a large gradient in S. To
avoid a significant bias in the SVD selection process we generate
curves of averaged Strehl ratio and Strehl $rms$ on concentric
annuli up to a maximum outer radius of about 85\% of the whole reconstructed
FoV radius. Once we have generated the above sets of curves for each
GS configuration and $DM_{2}$ conjugation altitude the SVD threshold
value selection is based on (i) having flat curves of averaged Strehl
within concentric annuli (in general we observe curves which are mostly
monotonically increasing or decreasing, so that we select those curves
with the lowest peak-to-valley values and/or the lowest gradient on
annulus averaged Strehl), (ii)~having the smallest Strehl $rms$
and (iii) in those situations where several threshold values yield
similar results based on the previous two points we select the largest
SVD threshold value. This is justified in terms of SNR considerations.

\subsection{Performance using Natural Guide Stars\label{sec:results_NGS}}

\begin{figure*}
\centering
  \includegraphics[width=17cm]{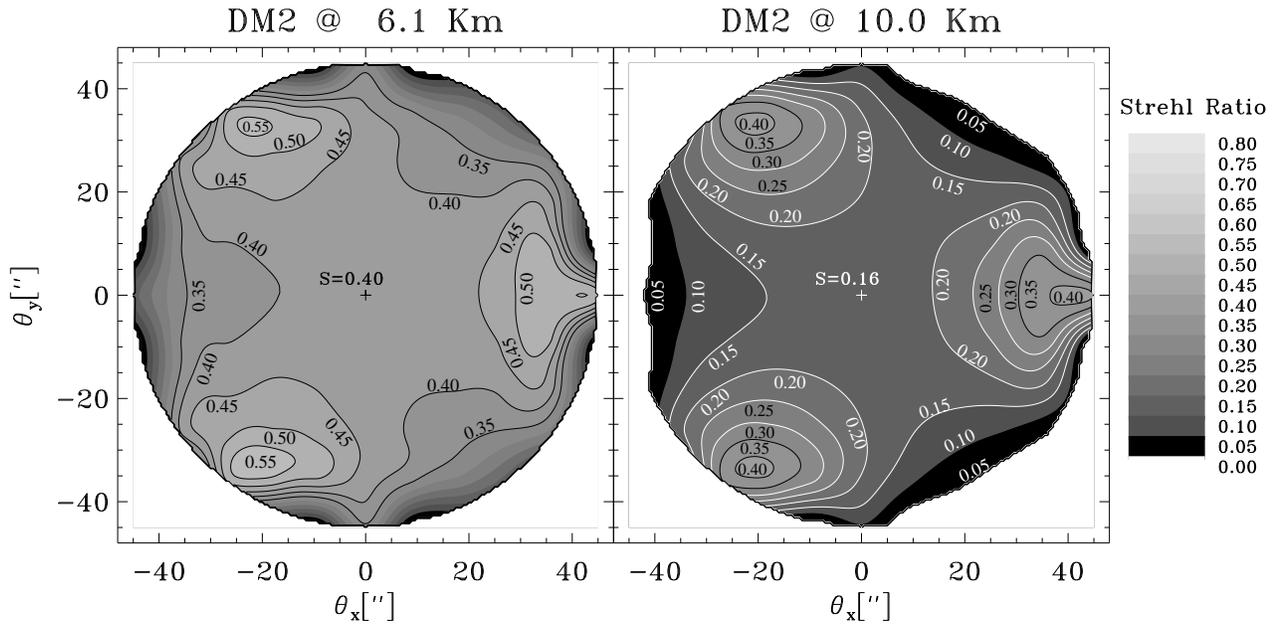}
  \caption{Strehl maps for a reconstructed $1.5\arcmin $ FoV for $DM_{2}$
           conjugated to 6.1~Km and 10.~Km for a constellation of 3NGS on a
           circumference of $45\arcsec $ radius. These maps correspond to the
           best SVD case for each $DM_{2}$ conjugation
           altitude.}
  \label{fig:iso_3NGS_ABP_45arc_DM2_06_10Km}
\end{figure*}

\begin{figure*}
  \centering
  \includegraphics[height=17cm,angle=90]{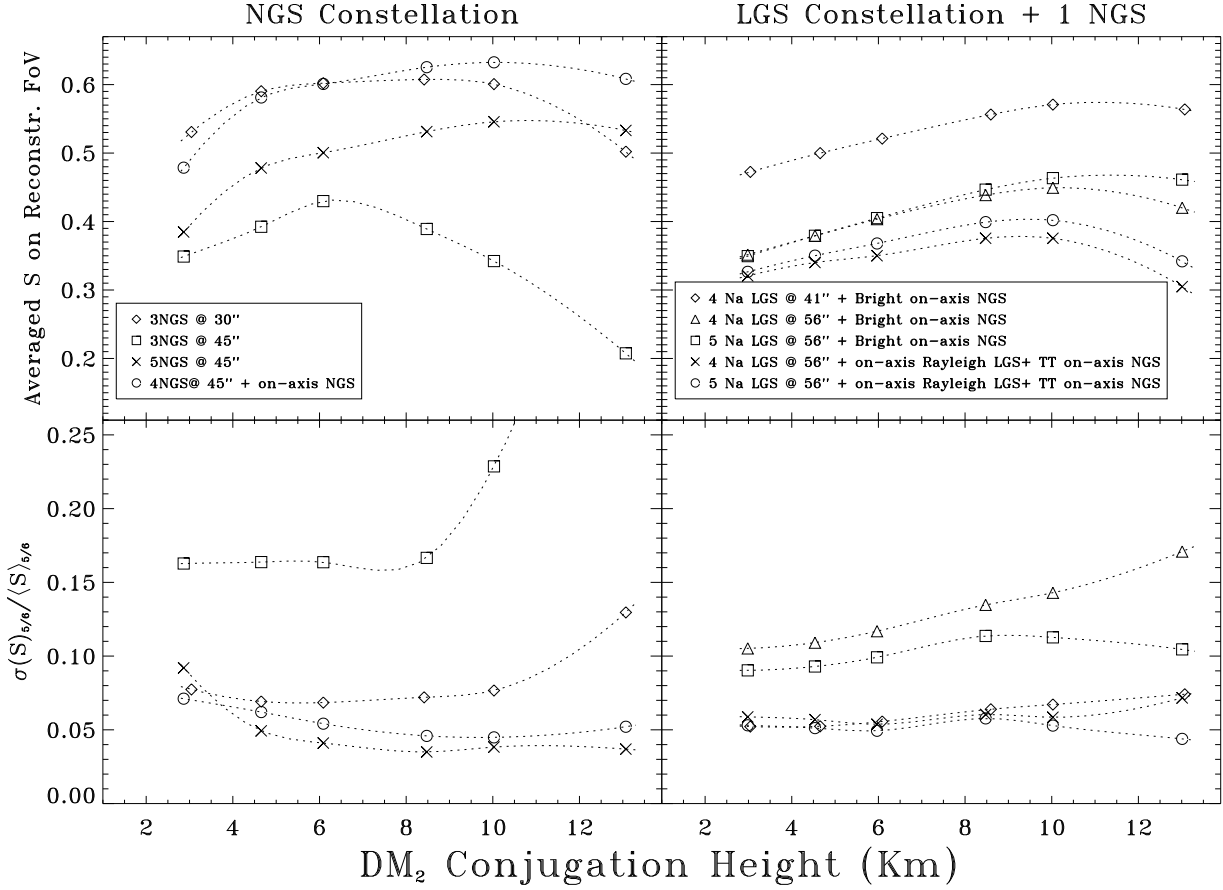}
  \caption{Summary of simulation results. Top panels show the dependence of the
           MCAO average Strehl ratio on the conjugation altitude of $DM_{2}$
           . The bottom panels give the normalized $rms$ of the averaged Strehl
           ratio on the 83\% radius annulus (see main text in
           Sect.~\ref{sec:results} for further
           details).}
  \label{fig:summary_Results_Rayleigh}
\end{figure*}

\begin{figure*}
  \centering
  \includegraphics[width=17cm]{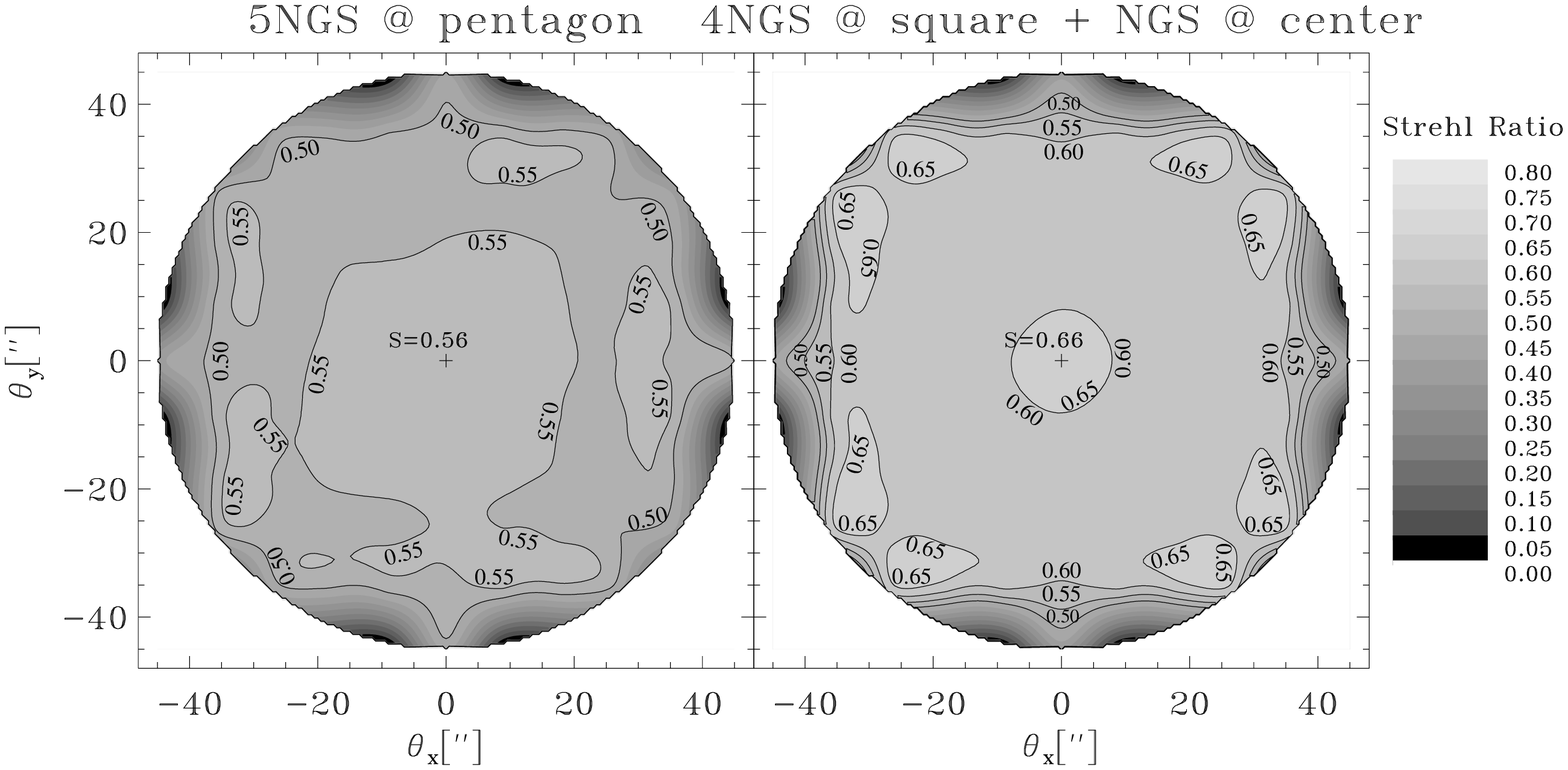}
  \caption{Checking two constellation geometries employing 5 NGS. On the
           \textbf{left panel} the Strehl map when the 5NGS are evenly arranged
           on the vertices of a regular pentagon within a $1.5\arcmin $ FoV. On
           the \textbf{right panel} the results when 4 NGSs are placed on the
           vertices of a square and 1 NGS is placed on the FoV center. Both
           panels correspond to having $DM_{2}$ conjugated to 10~Km above the
           telescope pupil.}
  \label{fig:iso_5NGS_4NGS_ABP_45arc_DM2_10Km}
\end{figure*}

Fig.~\ref{fig:iso_3NGS_ABP_30arc_DM2_08Km} shows K-band Strehl maps for the case
of having three natural guide stars evenly spaced on a circumference of radius
$30\arcsec $ in order to reconstruct a circular field of view of $1\arcmin $
diameter for several SVD threshold values. The panels correspond to setting
different singular value thresholds; it can be seen that the performance is
fairly constant until a large number (15) of SVD modes are rejected, some of
them with important contributions of absolute Zernike tilts on each of the two
reconstructed layers. When only a few SVD modes are rejected (i.e. from
THR=$7\times 10^{-5}$ to THR=$3\times 10^{-4}$ involving 2 to 7 SVD modes
rejected) the Strehl ratio peaks at the positions of the guide stars as would be
expected considering the least-squares reconstruction process employed while the
Strehl ratio across the reconstructed FoV is relatively uniform when rejecting 8
to 10 SVD modes (THR=$4\times 10^{-4}$ to THR=$10^{-3}$). This is a general
behavior observed in our simulations: uniformity of the Strehl ratio across the
reconstructed FoV is achieved at the cost of lowering the peak Strehl ratio at
the GS positions by an adequate SVD
filtering. Fig.~\ref{fig:iso_3NGS_ABP_45arc_DM2_06_10Km} shows the best SVD
Strehl maps when the three natural guide stars are on a circumference of radius
of $45\arcsec $, corresponding to a reconstructed FoV of $1.5\arcmin $ diameter,
and $DM_{2}$ conjugated to 6.1 and 10 Km above the telescope pupil. We observe
that with $DM_{2}$ conjugated to 10~Km the average Strehl ratio is significantly
smaller than when $DM_{2}$ is conjugated to 6.1~Km while the variation of the
Strehl ratio across the field is more pronounced for $DM_{2}$ conjugated to
10~Km. Making measurements on the maps it is found that the average Strehl ratio
drops from 0.43 to 0.20 and the normalised Strehl $rms$ increases from 16\% to
48\% (see also Fig.~\ref{fig:summary_Results_Rayleigh}).  The Strehl $rms$ value
when $DM_{2}$ is conjugated to 10~Km (and to 13~Km) is due to such a strong
center-to-edge Strehl variation that there is little sense in talking of an
average reconstructed Strehl ratio for this scenario.

By varying the $DM_{2}$ conjugation altitude in our simulations we
obtained that for constellations of 3 NGS the optimal conjugation
altitudes are approximately 8.5~Km and 6~Km for reconstructed FoVs
of $1\arcmin $ diameter and $1.5\arcmin $ diameter, respectively.
These optimal altitudes are surprisingly low when we consider the
vertical distribution of turbulence in the 7-layer model in Table~\ref{tab:averballprofile}.
We would expect the pupil-conjugate DM to correct the turbulence in
the four lower layers (up to 5~Km), and the second DM would therefore
be positioned to correct the turbulence in the three upper layers
(10-18~Km). \citet{TokovininA:limpt} published expressions allowing
optimal conjugation altitudes to be calculated for up to three mirrors
for a given turbulence profile. Applying these expressions to our
case (i.e. two DMs, balloon 7-layer model), we obtain an optimal $DM_{2}$
altitude of 13~Km. We believe the reason for such a disagreement
has to do with the strong assumptions in the aforesaid theoretical
work: the authors assume that whatever the altitudes of conjugation
for the DMs, tomography is able to perfectly reconstruct the wavefront
at those altitudes so that the only effect considered is the correction
of a continuous turbulence profile by a finite number of $DM$s. In
fact, the further from the telescope we conjugate $DM_{2}$, the less
overlap between the GS footprints at the metapupil at the $DM_{2}$
conjugation altitude. This effect will tend to push down the optimal
conjugation altitude for $DM_{2}$. This statement is supported by
the observation that the optimal conjugation altitude is lower for
the case of a FoV of $1.5\arcmin $ than for the case of a FoV of
$1\arcmin $. In order to further test this hypothesis, we carried
out simulations using five natural guide stars and keeping the rest
of the parameters of the simulation as in the case of 3 NGS. Increasing
the number of NGS in our constellation from 3 to 5 has the effect
of increasing the overlap between their footprints and thus we decrease
the importance of voids in the metapupil for the wavefront reconstruction
at that range. A comparison between the results achieved with constellations
of 3 NGS and constellations of 5 NGS is given in the left top and
left bottom panels in Fig.~\ref{fig:summary_Results_Rayleigh} where
we plot the average Strehl ratio on the reconstructed FoV as a function
of the $DM_{2}$ conjugation altitude and the Strehl ratio $rms$
variation, respectively (the right top and right bottom panels
of Fig.~\ref{fig:summary_Results_Rayleigh} show the results obtained
when considering different LGS constellations plus an on-axis NGS
and which will be the subject of the following sections). We can see
in Fig.~\ref{fig:summary_Results_Rayleigh} that when increasing
the number of NGSs from 3 to 5 NGS the conjugation altitude of $DM_{2}$
also increases and this effect can only be due to a better sampling
(and reconstruction) of the turbulence at the height at which $DM_{2}$
is conjugated. From these results we conclude that the effect neglected
by \citet{TokovininA:limpt} is very important when determining the
best $DM_{2}$ conjugation altitude.

\begin{figure*}
  \centering
  \includegraphics[width=17cm]{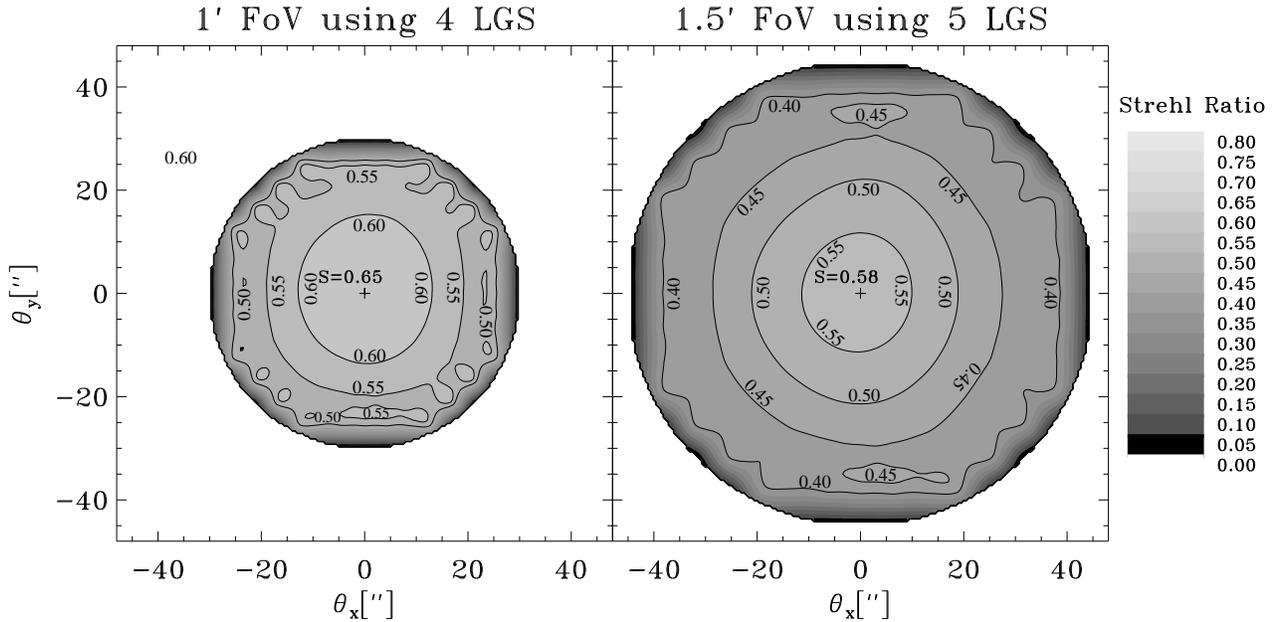}
  \caption{\textbf{Left panel} shows the best SVD Strehl map in a reconstructed
           FoV of $1\arcmin $ diameter with $DM_{2}$ conjugated to 10~Km and 4
           LGS on a circumference of $41.46\arcsec $ radius plus one on-axis
           bright NGS (i.e. wave-front sensing on the on-axis NGS is up
           to the same order as with any LGS). \textbf{Right panel} is
           the Strehl map corresponding to the reconstruction of a $1.5\arcmin $
           diameter FoV with $DM_{2}$ conjugated to 10~Km and 5 LGS on a
           circumference of $56.46\arcsec $ radius plus one on-axis bright
           NGS.}
  \label{fig:iso_4LGS_5LGS}
\end{figure*}

We also compared two different geometries employing 5 NGS: placing
all NGS on the vertices of a pentagon or putting 4 of them on the
corners of a square and one at the center. It turns out that it is
the latter configuration which yields better results (see top left
panel in Fig.~\ref{fig:summary_Results_Rayleigh} and iso-Strehl
contour plots in Fig.~\ref{fig:iso_5NGS_4NGS_ABP_45arc_DM2_10Km})
despite the fact that it leaves unsampled a slightly bigger portion
of the metapupil. This suggests that not only the fractional sampling
of the metapupil is important but also the distance of points within
the reconstructed FoV to the GS. Although we have not conducted any
test to check our explanation of this effect we believe that this
may be due to the LSE reconstruction in which by definition the best
reconstructed positions are those corresponding to the GS positions
while for a MAP reconstructor we may encounter a different behavior
since it is built by optimizing the reconstruction at a set of positions
which do not have to be coincident with those of the GS.

\subsection{Performance using Sodium Laser Guide Stars\label{sec:results_LGS}}

The probability of finding constellations of 3 (not to mention 5) bright natural
guide stars arranged to nicely span fields of $1\, -\, 1.5\, \arcmin $ is very
low. As an indication, using the Guide Star Catalogue-II\footnote{Copyright
Association of Universities for Research in Astronomy, Inc. Further details and
access to database at http://www-gsss.stsci.edu/gsc/gsc2/gsc22\_release\_notes.htm} we estimate that
the probability of finding three stars brighter than $m_\mathrm{R}=16$, which is
an optimistic value for limiting guide star magnitude, is approximately 3\% for
a field of $2\arcmin $ at galactic coordinates $(l,b)\, =\, (90\degr \, ,40\degr
)$, and this value will be further reduced when requiring the 3 NGS to be
properly arranged. In view of these considerations it will be necessary to
employ laser guide stars (LGS) in order to provide reasonable sky coverage for
star-oriented MCAO systems, at least on 8-10m telescopes.

It has already been pointed out by different authors \citep[see
e.g.][]{EllerbroekB:metct} that the indetermination of the tip-tilt of
individual LGS leads to tip-tilt anisoplanatism in an LGS-based MCAO system. The
tip-tilt anisoplanatism may be solved in a natural way by detecting the tip-tilt
of several NGS in the FoV \citep{RigautF:prilp}. In the Gemini-South MCAO system
it is planned to reconstruct the tip-tilt using three dim
(i.e. $m_\mathrm{R}\sim 19-20$) natural guide stars, and the sky coverage in the
K band is predicted \citep{RigautF:prilp} to be 15\% and 80\% at Galactic pole
and $b=30\degr $, respectively.  Alternative schemes can be derived by
considering that tip-tilt anisoplanatism mainly arises from turbulence-induced
quadratic wavefront errors (i.e defocus and astigmatism) at altitudes which
cannot be determined from the multiple-LGS measurements. If quadratic
measurements are made simultaneously with guide stars at different ranges, then
it may be possible to correct the tip-tilt anisoplanatism. One approach along
these lines is to make wavefront measurements on a natural guide star
simultaneously with the laser guide stars. We determine the performance of such
an approach here for fields of view of diameter $1\arcmin $ and $1.5\arcmin
$. In the case of the smaller field of view we employ four Na LGS on a
circumference spanning the field of view plus one on-axis NGS. In the case of
the $1.5\arcmin $ field of view we consider two Na LGS configurations; (i) four
LGS spanning the field of view plus one on-axis NGS (ii) five LGS on a
circumference spanning the field of view plus one on-axis NGS. The
circumferences on which the LGS are placed are somewhat larger than the fields
of view so that the metapupil sizes are the same when using cone shaped beams
from the LGS as when using cylindrical-shaped beams from NGS placed at the edge
of the field of view. Fig.~\ref{fig:iso_4LGS_5LGS} shows the best Strehl maps
for the $1\arcmin $ FoV (left panel) and $1.5\arcmin $ FoV (right panel),
employing constellations of 4 off-axis LGS+ on-axis NGS and 5 off-axis LGS +
on-axis NGS, respectively . For the $1\arcmin $ FoV case the average Strehl
ratio is similar to that obtained on the same field of view using 3 NGS although
it exhibits a larger peak-to-valley value. The $rms$ Strehl variation is similar
to the 3 NGS case but it does not rise steeply when $DM_{2}$ is conjugate to an
altitude higher than the optimal which for the LGS case is approximately 10~Km
while in the case of NGS constellation it was about 8.5~Km. In the case of the
$1.5\arcmin $ FoV the average Strehl ratio is almost the same for both 4LGS and
5LGS configurations. The optimal altitude occurs in the range 8-10~Km. If
$DM_{2}$ is conjugate to an altitude higher than the optimal then the average
Strehl ratio is higher and the rms variation is lower when 5 LGS are employed
rather than 4 LGS.

\subsection{Hybrid Rayleigh-Na system performance\label{sec:results_hybrid}}

\begin{figure*}
  \centering
  \includegraphics[width=17cm]{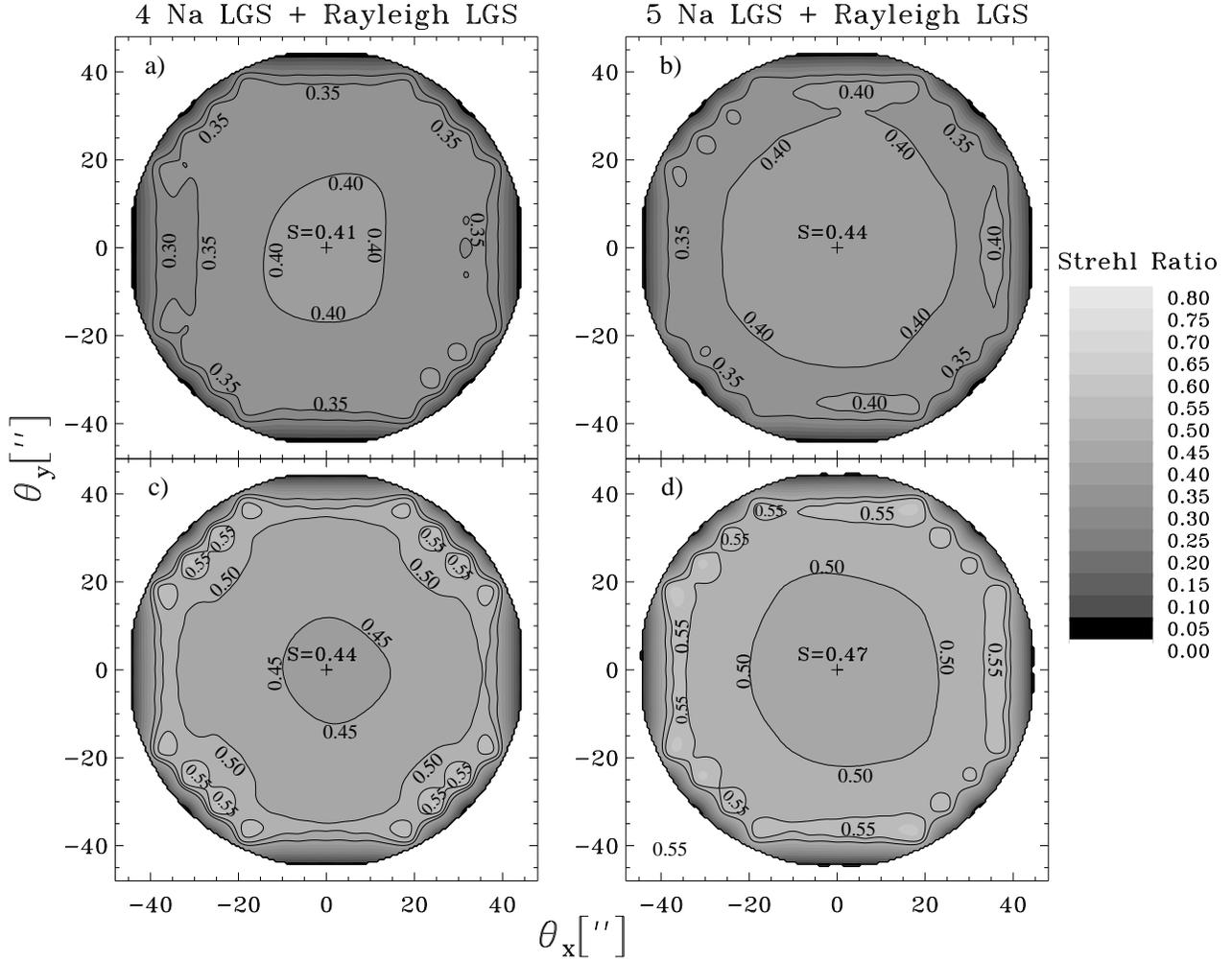}
  \caption{Strehl maps for two LGS geometries employing the hybrid LGS
           constellations.  \textbf{Panel a)} corresponds to using 4 Na LGS at
           90 Km with an off-axis angle of $56.46\arcsec $ (in order to provide
           a corrected $1.5\arcmin $ FoV) and an on-axis Tip-Tilt
           NGS. \textbf{Panel b)} corresponds to using 5 Na LGS at 90 Km with an
           off-axis angle of $56.46\arcsec $ plus an on-axis Rayleigh LGS at
           30~Km and an on-axis Tip-Tilt NGS.  Both cases consider $DM_{2}$
           conjugated to 10~Km. \textbf{Panels c)} and \textbf{d)} are the the
           same as panels a) and b), respectively except that tip-tilt is
           perfectly removed. This ideal exercise supports our interpretation on
           the limited level of reconstruction being mainly caused by residual
           tip-tilt modes.}
  \label{fig:iso_hybridLGS}
\end{figure*}

The natural guide star employed to measure quadratic modes has to be bright,
thereby again compromising sky coverage. An alternative approach to solving the
tip-tilt anisoplanatism caused by the quadratic modes consists of employing both
Rayleigh and Na LGS as well as a single natural guide star which is only used to
determine global tip-tilt
\citep{FemeniaB:numsm,EllerbroekB:metct,FemeniaB:larss}. Here we apply the
algorithm described in \citet{FemeniaB:larss} whose starting point is to realize
that unresolved second order modes in LGS-based tomography can be reconstructed
using an LGS constellation made at several altitudes (e.g. Rayleigh + Sodium
LGSs, or Rayleigh LGSs at different altitudes) as shown by
\citet{BrusaG:multAO}. The reconstruction process consists of two stages in
which barycenter tip-tilt is measured from a NGS and in parallel we perform
tomography with the LGS constellation placed at at least two different
altitudes. From the LGS tomography we obtain a reconstruction of all modes
starting from the Zernike second-order modes and this information is used
together with the NGS barycenter tip-tilt to obtain an estimate of the global
Zernike tilts in the entire reconstructed FoV. The hybrid technique may be
implemented in different ways, in particular \citet{FemeniaB:larss} considers a
single on-axis Rayleigh LGS on which to conduct high-order wave-front sensing
while \citet{EllerbroekB:metct} consider several off-axis Rayleigh LGS coupled
to low-order WFS. Other parameters to be optimized include the height of the
Rayleigh LGS and the wavelength for the Rayleigh LGS. Given the number of new
parameters to be investigated the results we present here are by no means
exhaustive but only preliminary to give an idea of the potential of the
technique. We assume two configurations of 4~Na~LGS and 5~Na~LGS evenly arranged
on a circumference of $56.45\arcsec $ radius (so that the reconstructed FoV is
$1.5\arcmin$) plus an on-axis Rayleigh LGS. The Na LGSs are observed at $589\
\mathrm{nm}$ and the Rayleigh LGS at $355\ \mathrm{nm}$ with $16\times 16$
subaperture SHSs. The rest of the simulation parameters (atmospheric profile,
number of CCD pixels and SHS FoV, no noise in the SHS, number of reconstructed
modes, etc.) are the same as in the simulations with NGS and LGS+on-axis
NGS. The top panels a) \& b) in Fig.~\ref{fig:iso_hybridLGS} shows the
reconstruction across the $1.5\arcmin $ FoV for the cases of 4~Na~LGS and
5~Na~LGS, respectively.

On comparing the results from the constellation using 4 Na LGS against
the results using 5 Na LGS we see a limited gain. This limited performance
gain is a consequence of the limited reconstruction along the on-axis
direction which in turn is used to reconstruct the Zernike tip-tilt
modes across the entire reconstructed FoV. This statement is supported
by the levels of reconstruction attained if we assume perfect reconstruction
of the Zernike tip-tilt modes across the entire FoV as depicted in
panel c) in Fig.~\ref{fig:iso_hybridLGS} for the case of 4 Na LGS
at $56\arcsec $ and in panel d) in Fig.~\ref{fig:iso_hybridLGS}
for the case of 5 Na LGS at $56\arcsec $; in both cases also considering
on-axis Rayleigh LGS at 30 Km and on-axis Tip-tilt NGS. When comparing
in Fig.~\ref{fig:iso_hybridLGS} the bottom panels (perfect correction
of Zernike tip-tilt across the entire FoV) against the top panels
(actual correction of Zernike tilts) we verify that in the latter
figures there is a small amount of tilt anisoplanatism caused by a
worse correction of higher-order modes (especially second-order modes)
in the on-axis direction. This in turn translates into tilt anisoplanatism
when using those higher-order modes to reconstruct off-axis Zernike
tilts. This situation also appeared when dealing with 5 NGS where
we observed that the best configuration is that corresponding to 4
NGS on the corners of a square plus one on-axis NGS rather than all
5 NGS on the vertices of a pentagon. Based on this conjecture it would
be more effective having 4 Na LGS distributed on the vertices of a
square plus one on-axis Na LGS than having all 5 Na LGS distributed
on the vertices of a pentagon.

\section{Choosing the $DM_{2}$ conjugation altitude \label{sec:choosing}}

As seen in Fig.~\ref{fig:summary_Results_Rayleigh} the optimal $DM_{2}$conjugation
altitude depends very strongly on the GS nature and configuration.
In the case of observing 3NGS on the vertices of equilateral triangles
on a circumference of radius $30\arcsec $ or $45\arcsec $ the optimal
altitudes are about 8.5 and 6.0 Km, respectively. When considering
5 NGSs the optimal $DM_{2}$ conjugation altitude is 10 Km. When hybrid
LGS constellations and a single tip-tilt NGS are used aiming at a
reconstructed FoV of $1.5\arcmin $ the best $DM_{2}$ conjugation
altitude occurs in the range {[}8,11{]} Km with little variation of
performance within that range leaving the final decision to opto-mechanical
considerations. In order to find an optimal $DM_{2}$ conjugation
altitude that best suits all possible GS configurations cases in this
work we consider the following figure of merit:

\[
\chi ^{2}(h)\, =\, \sum _{i=1}^{9}\left(\frac{S_\mathrm{max,i}-S_\mathrm{i}(h)}{S_\mathrm{max,i}}\right)^{2}\cdot w_\mathrm{i},\]
 where $h$ denotes a generic $DM_{2}$ conjugation altitude, the
index $\mathrm{i=1,\ldots ,9}$ runs over all the GS configurations in this
work, $w_\mathrm{i}$ weights different GS configurations, $S_\mathrm{max,i}$
is the maximum Strehl ratio achieved with the $i$th GS configuration
and $S_\mathrm{i}(h)$ is the achieved Strehl ratio with the $i$th GS configuration
when conjugating $DM_{2}$ to altitude $h$ above the telescope entrance
pupil. As the $\chi ^{2}$ mimization process will depend
on $\left\{ w_\mathrm{i}\right\} $ we have considered the sets of weights
listed in Table~\ref{tab:w-sets}: the first weighting set (i.e.
column $w_{1}$in Table~\ref{tab:w-sets}) gives equal importance
to all GS configurations, weighting set $\left\{ w_{2}\right\} $
is focused on avoiding system complexity and thus assigns higher weight
to simpler systems while weighting sets $\left\{ w_{3}\right\} $
and $\left\{ w_{4}\right\} $ are derived from simple sky coverage
considerations based on star counts from the Guide Star Catalogue-2
at $(l,b)\, =\, (90\degr \, ,40\degr )$ and assuming that for the
NGS we require at least $m_\mathrm{R}\lesssim 16$ when it is used for high-order
mode reconstruction and $m_\mathrm{R}\lesssim 18.5$ when it is only used
for tip-tilt reconstruction (i.e. as in the case of the hybrid laser
configurations). The difference between weighting sets $\left\{ w_{3}\right\} $
and $\left\{ w_{4}\right\} $ is that the former is proportional to
the probability of finding the required number of NGSs in the GS configuration
and the second is proportional to the square root of that probability
(i.e. $w_\mathrm{4,i}=\sqrt{w_\mathrm{3,i}}$ where index $i$ runs over the different
GS configurations under study). This choice for $\left\{ w_{3}\right\} $and
$\left\{ w_{4}\right\} $ is very qualitative but given the small
difference between the results obtained with them (see Table~\ref{tab:best-h})
we are confident that a detailed sky coverage analysis is likely to
differ little from the results obtained here. The $\chi ^{2}(h)$
minimization process yields an {}``optimal'' $DM_{2}$ conjugation
altitude as a function of the weighting set. The first and second
columns in Table~\ref{tab:best-h} give the $DM_{2}$ conjugation
altitude for which a given GS configuration attains its maximum Strehl
ratio and the Strehl ratio itself. Columns 3 to 6 give Strehl ratios
for each GS configuration when the $DM_{2}$ is conjugated to the
joint optimal conjugation altitude according to the weighing set under
consideration (i.e $S_\mathrm{opt,1}$corresponds to using $w_{1}$); the
value of such optimal conjugation altitude for each weighting set
being displayed in parenthesis below each column title. The last two
columns in Table~\ref{tab:best-h} give the Strehl ratio and fractional
decrease with respect to $S_\mathrm{max}$, respectively, when $DM_{2}$
is conjugated to 8~Km. 

\begin{table}

\caption{Different normalized-to-maximum weighting sets used to obtain the
$DM_{2}$conjugation altitude. Values marked with superindex $^{\mathrm{a}}$
should be further reduced when requiring proper arrangement of the
NGSs.\label{tab:w-sets}}

\begin{tabular}{|c|c|c|c|c|}
\multicolumn{1}{c}{GS Configuration}&
\multicolumn{1}{c}{$w_{1}$}&
\multicolumn{1}{c}{$w_{2}$}&
\multicolumn{1}{c}{$w_{3}$}&
\multicolumn{1}{c}{$w_{4}$}\\
\hline
\hline 
3 NGS @ $30\arcsec $&
1.00&
1.00&
$9\times 10^{-4}$$^{\mathrm{a}}$&
0.03$^{\mathrm{a}}$\\
\hline 
3 NGS @ $45\arcsec $&
1.00&
1.00&
$9\times 10^{-3}$$^{\mathrm{a}}$&
0.1$^{\mathrm{a}}$\\
\hline 
5 NGS @ $45\arcsec $&
1.00&
0.60&
$6\times 10^{-5}$$^{\mathrm{a}}$&
0.008$^{\mathrm{a}}$\\
\hline 
4 NGS @ $45\arcsec $ + NGS&
1.00&
0.60&
$6\times 10^{-5}$$^{\mathrm{a}}$&
0.008$^{\mathrm{a}}$\\
\hline 
4 LGS @ $41\arcsec $ + NGS&
1.00&
0.40&
0.22&
0.47\\
\hline 
4 LGS @ $56\arcsec $ + NGS&
1.00&
0.40&
0.44&
0.67\\
\hline 
5 LGS @ $56\arcsec $ + NGS&
1.00&
0.30&
0.44&
0.67\\
\hline 
Hybrid: 4 Na LGS &
1.00&
0.20&
1.00&
1.00\\
\hline 
Hybrid: 5 Na LGS &
1.00&
0.15&
1.00&
1.00\\
\hline
\end{tabular}
\end{table}

\begin{table*}

\caption{Finding the optimal DM2 conjugation altitude.\label{tab:best-h}}

\begin{tabular}{|c||cc||c|c|c|c||cc|}
\multicolumn{1}{c}{}&
\multicolumn{1}{c}{}&
\multicolumn{1}{c}{}&
\multicolumn{1}{c}{$S_\mathrm{opt,1}$}&
\multicolumn{1}{c}{$S_\mathrm{opt,2}$}&
\multicolumn{1}{c}{$S_\mathrm{opt,3}$}&
\multicolumn{1}{c}{$S_\mathrm{opt,4}$}&
\multicolumn{1}{c}{$S$}&
\multicolumn{1}{c}{$\varepsilon \, (\%)$}\\
\multicolumn{1}{c}{GS Configuration}&
\multicolumn{1}{c}{$h_\mathrm{max}$}&
\multicolumn{1}{c}{$S_\mathrm{max}$}&
\multicolumn{1}{c}{(7.7 Km)}&
\multicolumn{1}{c}{(6.7 Km)}&
\multicolumn{1}{c}{(8.4 Km)}&
\multicolumn{1}{c}{(9.0 Km)}&
\multicolumn{2}{c}{(8.0 Km)}\\
\hline
\hline 
3 NGS @ $30\arcsec $&
8.7&
0.61&
0.61&
0.60&
0.61&
0.61&
0.61&
-0.1\\
\hline 
3 NGS @ $45\arcsec $&
6.0&
0.43&
0.37&
0.41&
0.35&
0.30&
0.36&
-15.5\\
\hline 
5 NGS @ $45\arcsec $&
10.8&
0.55&
0.52&
0.51&
0.53&
0.54&
0.53&
-4.1\\
\hline 
4 NGS @ $45\arcsec $ + NGS&
10.1&
0.63&
0.62&
0.61&
0.63&
0.63&
0.62&
-1.8\\
\hline 
4 LGS @ $41\arcsec $ + NGS&
11.1&
0.57&
0.54&
0.53&
0.55&
0.56&
0.59&
-4.5\\
\hline 
4 LGS @ $56\arcsec $ + NGS&
10.2&
0.45&
0.43&
0.41&
0.44&
0.44&
0.43&
-3.6\\
\hline 
5 LGS @ $56\arcsec $ + NGS&
11.4&
0.47&
0.43&
0.42&
0.45&
0.45&
0.44&
-6.0\\
\hline 
Hybrid: 4 Na LGS &
9.3&
0.38&
0.37&
0.36&
0.38&
0.38&
0.37&
-1.8\\
\hline 
Hybrid: 5 Na LGS &
9.5&
0.40&
0.39&
0.38&
0.40&
0.40&
0.40&
-2.2\\
\hline
\end{tabular}
\end{table*}

\section{Conclusions\label{sec:conclusions}}

Modal MCAO in the {}``classical'' star-oriented fashion holds the
promise to reconstruct FoVs as large as $1.5\arcmin $ in diameter.
Ideally one would like to rely on NGS constellations but the chances
of finding a suitable one is extremely low even for the simplest case
of 3NGS. Thus if one wants to build a star-oriented MCAO system which
can be used over a significant fraction of the sky one must necessarily
consider LGS-based MCAO systems. However, one then has to face the
tip-tilt problem and the indetermination of second-order modes. The
simplest approach to LGS-based MCAO systems is that of relying on
a bright enough NGS on which one could perform not only tip-tilt \&
second-order sensing but also high-order WFS. This yields poor results
and the initially surprising result that performance does not improve
significantly when increasing the number of Na LGSs from 4 to 5. In
any case, whether we use 4 Na LGS or 5 Na LGS the reconstructed FoV
is strongly affected by tip-tilt anisoplanatism and the reconstruction
is extremely sensitive to the SVD filtering process which in practice
will be determined by the SNR in the WFS measurements. We also remark
that such a naive approach also has small sky coverage as it requires
a star which is bright enough to obtain high-order wavefront measurements.
However, with hybrid LGS constellations (Sect.~\ref{sec:results_hybrid})
it may be possible to obtain good levels of reconstruction. Taking
the sky coverage values obtained with the three tip-tilt NGS scheme
considered by Gemini (see Sect.~\ref{sec:results_LGS}) and assuming
equivalent magnitude requirements for the tip-tilt NGS with hybrid
LGS constellations it is expected that this approach can be applied
to a larger fraction of the sky as only a single tip-tilt NGS is required.
There is a lot of room for possible improvements to the technique
used \citep{FemeniaB:larss}. 

The optimal $DM_{2}$ conjugation altitude depends strongly on the
guide star constellation used. When considering NGS constellations
our MCAO simulations give results in disagreement with the theoretical
work by \citet{TokovininA:limpt} which would place $DM_{2}$ at about
13~Km above telescope pupil \citep{DevaneyN:adawb}. The reason for
such disagreement is believed to be caused by the strong assumptions
in these theoretical works resulting in overestimating the $DM_{2}$
conjugation altitude. From our study based on detailed numerical simulations
we consider that a suitable $DM_{2}$ conjugation altitude for a 10-m
class telescope is around 8~Km and thus we are considering such an
option for the design of the AO system at the GTC telescope which
considers an initial AO system with a single corrector but upgradeable
to a dual-conjugate system. It has been argued that the aforementioned
theoretical works would be applicable to the case of extremely large
telescopes (30-100 m diameter) as the main limiting assumption (i.e.
full metapupil coverage by overlapping the GS beams at the reconstructed
layers) is relaxed and it should be possible to fulfill the requirement
of perfect wavefront reconstruction at each metapupil. However, one
should consider that the larger the telescope diameter, the larger
the number of modes required in order to reach the same Strehl Ratio
as with a smaller telescope diameter. It follows from this
consideration that with extremely large telescopes, although the metapupil
coverage is nearly complete, the importance of small gaps in the metapupil
coverage is much more important, this implying that the assumptions
in the above theoretical works may not be realistic even for extremely
large telescopes.

Our last conclusion regards a rule of thumb indicated by our simulations:
given a fixed number of GS (either LGS or NGS) the best MCAO
performance occurs for the constellation configuration that minimizes
the distance from any point in the FoV to a GS position. As we have
seen although 5 GS on the vertices of a pentagon give a larger covered
surface of the metapupil than 4 GS on the square vertices plus one
on-axis GS, the latter configuration yields the better reconstruction.

We would like to end with a comment on the turbulence profile used
in our simulations. It has been derived from only 6 turbulence profiles
at the ORM obtained with balloon flights which is a rather small sample
of turbulence profiles. A larger set of turbulence profiles would
be desirable in order to obtain a more significant average profile
at ORM. In any case, all the simulations were also conducted on a
modified version of the average profile and the optimal conjugation
altitude of $DM_{2}$ remained essentially the same. 

\begin{acknowledgements}
The Guide Star Catalogue-II is a joint project of the Space
Telescope Science Institute and the Osservatorio Astronomico di Torino.
Space Telescope Science Institute is operated by the Association of
Universities for Research in Astronomy, for the National Aeronautics
and Space Administration under contract NAS5-26555. The participation
of the Osservatorio Astronomico di Torino is supported by the Italian
Council for Research in Astronomy. Additional support is provided
by European Southern Observatory, Space Telescope European Coordinating
Facility, the International GEMINI project and the European Space
Agency Astrophysics Division.
\end{acknowledgements}
\bibliographystyle{apj}
\bibliography{ao}

\end{document}